\documentclass[preprint2]{aastex}

\newcommand{\FUV}{{\it FUV~}}
\newcommand{\NUV}{{\it NUV~}}
\newcommand{\FUVnb}{{\it FUV}}
\newcommand{\NUVnb}{{\it NUV}}

\newcommand{\galex}{\textit{GALEX~}}

\shortauthors{}
\usepackage{natbib}
\bibpunct{(}{)}{;}{a}{}{,}

\begin{document}
\title{Clustering Properties of restframe UV selected galaxies II:\\
Migration of star formation sites with cosmic time from GALEX and
CFHTLS}

\author{ 
  S\'ebastien Heinis\altaffilmark{1,2},
  Bruno Milliard\altaffilmark{1}, 
  St\'ephane Arnouts\altaffilmark{1},
  J\'er\'emy Blaizot\altaffilmark{1,3},\\ 
  David Schiminovich\altaffilmark{4}, 
  Tam\'as Budav\'ari\altaffilmark{2},
  Olivier Ilbert\altaffilmark{5},
  Jos\'e Donas\altaffilmark{1}, 
  Marie Treyer\altaffilmark{1},\\ 
  Ted K. Wyder\altaffilmark{6},
  Henry J. McCracken\altaffilmark{7},
  Tom A. Barlow\altaffilmark{6}, 
  Karl Forster\altaffilmark{6},
  Peter G. Friedman\altaffilmark{6},\\
  D. Christopher Martin\altaffilmark{6},
  Patrick Morrissey\altaffilmark{6},
  Susan G. Neff\altaffilmark{8},
  Mark Seibert\altaffilmark{6},
  Todd Small\altaffilmark{6},
  Luciana Bianchi\altaffilmark{9},
  Timothy M. Heckman\altaffilmark{2},
  Young-Wook Lee\altaffilmark{10},
  Barry F. Madore\altaffilmark{11},\\
  R. Michael Rich\altaffilmark{12},
  Alex S. Szalay\altaffilmark{2},
  Barry Y. Welsh\altaffilmark{13},
  Sukyoung K. Yi\altaffilmark{10} and
  C. K. Xu \altaffilmark{6}
 }
\altaffiltext{1}{Laboratoire d'Astrophysique de Marseille, BP 8, Traverse
du Siphon, 13376 Marseille Cedex 12, France}

\altaffiltext{2}{Department of Physics and Astronomy, The Johns Hopkins
University, Homewood Campus, Baltimore, MD 21218}

\altaffiltext{3}{Max Planck Institut f\"ur astrophysik, D-85748
Garching, Germany}

\altaffiltext{4}{Department of Astronomy, Columbia University, New
York, NY 10027}

\altaffiltext{5}{Institute for Astronomy, 2680 Woodlawn Drive,
Honolulu, HI 96822}

\altaffiltext{6}{California Institute of Technology, MC 405-47, 1200 East
California Boulevard, Pasadena, CA 91125}

\altaffiltext{7}{Institut d'Astrophysique de Paris, Universit\'e
Pierre et Marie Curie, UMR 7095, 98 bis Bvd Arago, 75014 Paris,
France}

\altaffiltext{8}{Laboratory for Astronomy and Solar Physics, NASA Goddard
Space Flight Center, Greenbelt, MD 20771}

\altaffiltext{9}{Center for Astrophysical Sciences, The Johns Hopkins
University, 3400 N. Charles St., Baltimore, MD 21218}

\altaffiltext{10}{Center for Space Astrophysics, Yonsei University, Seoul
120-749, Korea}

\altaffiltext{11}{Observatories of the Carnegie Institution of Washington,
813 Santa Barbara St., Pasadena, CA 91101}

\altaffiltext{12}{Department of Physics and Astronomy, University of
California, Los Angeles, CA 90095}

\altaffiltext{13}{Space Sciences Laboratory, University of California at
Berkeley, 601 Campbell Hall, Berkeley, CA 94720}

\begin{abstract}
  {We analyze the clustering properties of ultraviolet selected
    galaxies by using \galex-~SDSS data at $z<0.6$ and CFHTLS deep
    $u'$ imaging at $z\sim 1$. These datasets provide a unique basis
    at $z\le 1$ which can be directly compared with high redshift
    samples built with similar selection criteria. We discuss the
    dependence of the correlation function parameters ($r_0$,
    $\delta$) on the ultraviolet luminosity as well as the linear bias
    evolution. We find that the bias parameter shows a gradual decline
    from high ($b_8 \gtrsim 2$) to low redshift ($b_8 \simeq
    0.79^{+0.1}_{-0.08}$). When accounting for the fraction of the
    star formation activity enclosed in the different samples, our
    results suggest that the bulk of star formation migrated from high
    mass dark matter halos at $z>2$ ($10^{12} \le M_{min} \le 10^{13}
    M_{\odot}$, located in high density regions), to less massive
    halos at low redshift ($M_{min} \le 10^{12} M_{\odot}$, located in
    low density regions). This result extends the ``downsizing''
    picture (shift of the star formation activity from high stellar
    mass systems at high $z$ to low stellar mass at low $z$) to the
    dark matter distribution.}\vspace{1.cm}
\end{abstract}
\shorttitle{Migration of star formation sites with cosmic time from \galex and
CFHTLS}
\slugcomment{Submitted for publication in the Special GALEX Ap.J.Suppl. Issue}
\maketitle

\section{Introduction}

Accumulated evidence shows that the cosmic Star Formation Rate (SFR)
has been decreasing from $z \sim 1$ by a dramatic factor of about $5$
\citep{Hopkins_2004, Lilly_1996, Madau_1996, Schiminovich_2005,
Sullivan_2000, Wilson_2002}. This is linked to the decrease of the
contribution of the faint galaxies that dominate the star formation
density, and to the strong decline of the most ultraviolet-luminous
galaxies with time, given the redshift evolution of the 1500 \AA~
luminosity function \citep{Arnouts_2005}. Another aspect of this
evolution, known as ``downsizing'' \citep{Cowie_1996}, is the
observation that star formation activity shifts with time from high to
low stellar mass systems \citep[][and references therein]{Bundy_2005,
Jimenez_2005, Juneau_2005, Heavens_2004}.

The star formation history results from the interplay between the
physical processes driving the star formation fueling (gas cooling)
and regulation (feedback), both closely related to galaxy
environment. Recent simulations show that about half of the galaxy gas
is accreted through a cold mode, which dominates at high redshift in
high density environments, and shifts to low density environments in
the local Universe \citep{Keres_2005}. The type of the dominant
feedback process is expected to depend on galaxy host halo mass:
supernovae explosions \citep[e.g.][]{Benson_2003} at low mass, and AGN
\citep[e.g.][]{Croton_2006} at high mass. \citet{Cattaneo_2006} show
that the introduction of a critical halo mass above which there is a
complete shutdown of cooling and star formation is efficient to
reproduce the bimodality in galaxy properties observed in the local
Universe \citep[e.g.][]{Baldry_2004}.

In this paper, we propose to set constraints on the roles of these
different processes through cosmic time by assessing the spatial
distribution of star formation in the Universe from high to low
redshifts. A convenient method is to study the clustering properties
of restframe ultraviolet (UV) selected galaxies. This has already been
performed at high redshifts using Lyman Break Galaxies (LBGs) samples
to show that, at these epochs, star formation is highly clustered and
concentrated in overdense regions \citep{Adelberger_2005, Allen_2005,
Arnouts_2002, Foucaud_2003, Giavalisco_2001}. The study of the
redshift evolution of the clustering properties of actively star
forming galaxies has now been made possible in a homogeneous way with
the combination of restframe UV data collected from $z\sim4$ to $z =
0$. To extend high-$z$ studies, we use GALEX observations in the
recent Universe and CFHTLS deep imaging at $z=1$. We compute the
angular correlation function (ACF) of star forming galaxies and derive
their bias and its evolution.

In a companion paper, \citet[][hereafter Paper I]{Milliard_2007}, we
describe in detail the methodology and the first results of the
angular correlation function measurements of UV selected galaxies
using a \galex sample at $z\le 0.6$. Section \ref{sec_sample}
summarizes the sample properties and presents a new restframe
UV-selected sample from the $u'$ band deep CFHTLS imaging survey that
we use to extend the analysis to higher redshift ($z\sim 1$). We then
investigate the dependence on redshift and UV luminosity of the
clustering properties: $r_0$, $\delta$ in sect. \ref{sec_results_Muv},
bias in sect. \ref{sec_bias}. In the last section we discuss the
evolution of the preferred sites of star formation over the last 90\%
of the age of the Universe.  

All magnitudes have been corrected for Galactic extinction using the
E(B-V) value from the \citet{Schlegel_1998} maps and the
\citet{Cardelli_1989} extinction law.  Throughout the paper, we adopt
the following cosmological parameters: $\Omega_{m}=0.3$,
$\Omega_{\Lambda}=0.7$, $H_0=70$ km s$^{-1}$Mpc$^{-1}$.

\section{Samples description}\label{sec_sample}
\subsection{\galex}\label{sec_galex_sample}

In this work, we use the same subsample of \galex Release 2 (GR2)
Medium Imaging Survey (MIS) fields cross-matched with SDSS DR5
presented in Paper I, and we refer to this paper for a full
description. We recall here the main characteristics of the
selection. We only keep \galex objects with SDSS counterparts within a
search radius of 4\arcsec and use the closest SDSS match. We select
galaxies as objects with SDSS \texttt{type} equal to 3. We use the
half of the MIS fields from the GR2 dataset with the lowest Galactic
extinction $\left( \langle E(B-V)\rangle \leq
0.04\right)$. Photometric redshifts are computed using an empirical
method \citep{Connolly_1995} trained on SDSS spectroscopic
counterparts. The standard deviation estimated from the SDSS
spectroscopic redshifts is $\sigma = 0.03$. We then use a template
fitting procedure \citep{Arnouts_2070} to derive UV luminosities. Our
starting samples include objects with \FUV $<22$ or \NUV $<22$.

\begin{figure}[!t]
  \plotone{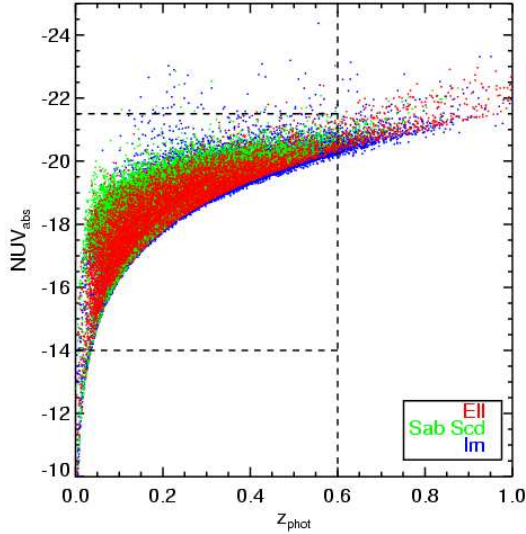}
  \caption{ \small \NUV absolute magnitude-photometric redshift
  relation in the \galex sample. The colors code the type according to
  the best fitting template: red represent elliptical types, green
  spirals, and blue irregulars. The dashed lines indicate the additional
  cuts adopted: $z_{phot}<0.6$ and $-21.5<NUV_{\rm{abs}}<-14.$ The same cuts
  hold for \FUVnb.}  \label{fig_Mabs_zp}
\end{figure}

The \NUV absolute magnitude vs photometric redshift relation is shown
in figure \ref{fig_Mabs_zp}. The colors code the galaxy type
determined using a SED template fitting procedure: red represent
elliptical types, green spiral and blue irregular. We restricted
hereafter the samples to $-21.5<NUV_{\rm{abs}}<-14.$ and
$0.<z_{phot}<0.6$ (dashed lines on fig \ref{fig_Mabs_zp}). The same
cuts have been applied to the \FUV sample.

\begin{figure}[!t]
  \plotone{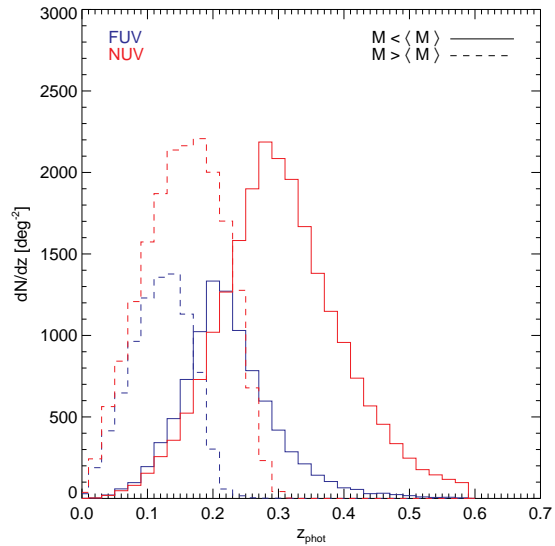}
  \caption{\small Redshift distribution of the subsamples cut in
  absolute UV magnitude: $M<\langle M \rangle$, solid lines;
  $M>\langle M \rangle$, dashed lines, \FUV is shown as blue and \NUV
  as red.}  
  \label{fig_Muv_nzs}
\end{figure}

In the following, we consider both \FUV and \NUV bands and we divide
the samples in two bins according to the mean $UV$ absolute
magnitude. The figure \ref{fig_Muv_nzs} shows the photometric redshift
distributions of the \galex samples; the table \ref{tab_Muv_galex}
summarizes the properties of the samples.

\subsection{CFHTLS}\label{sec_cfhtls_sample}

 The CFHTLS-Deep survey consists of deep multi-colour images collected
through the $u' g'r' i' z'$ filters over four independent areas of 1
$deg^2$ each and reaching the limiting magnitude of $i'_{AB}\sim 26$.
In this work, we use the official CFHTLS data release T0003. For a
full presentation of the CFHTLS-Deep survey, we refer to
\citet{Schultheis_2006}\footnote{see also
\url{http://www.cfht.hawaii.edu/Science/CFHLS/} and
\url{http://www.ast.obs-mip.fr/article204.html}}. We built specific
masks from the $u$-band images to mask out stars, chips edges' and
artifacts. The total solid angle of the four fields used after masking
is 3.1 $deg^2$. The star/galaxy separation is based on the same method
as \citet{McCracken_2003} with the half-light radius versus $u$
magnitude plot. This selection has been applied down to $u =
23$. Beyond this limit, we combine the photometric criterion with the
star/galaxy classification derived from the photometric redshift code
{\it Le Phare}, \citep{Arnouts_2070}.  \\
 
\begin{figure}[t]
  \plotone{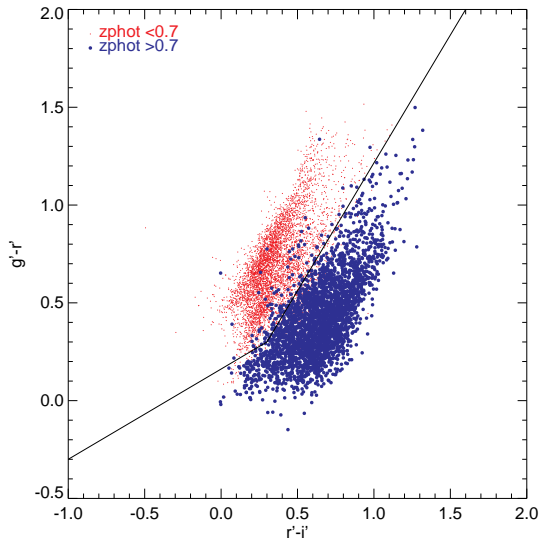}
  \caption{\small Redshift selection based on the ($g'-r'$) vs
    ($r'-i'$) diagram for the CFHTLS-D1 field.  The small dots show
    galaxies with $z_{phot}<0.7$ and big dots galaxies with
    $z_{phot}>0.7$.  The line represents the adopted color-color
    selection criterion.}

  \label{fig_color_sel} 
\end{figure}

To construct the sample of UV selected galaxies at $z\sim 1$, we adopt
a $u'$ magnitude limit of $u'=24$, which ensures a genuine
$UV$-selected sample as the $u'$ effective wavelength (3587 \AA)
corresponds to 1848 \AA~at our mean redshift $\langle z \rangle =
0.94$.  The fraction of objects lost (without any redshift selection)
with a $i'=26$ cut is on average 0.07\% over the four fields at
$u'=24$. The redshift selection of the sample is based first on a
color-color selection and then on the photometric redshift
selection. We do not adopt a single selection based on the photometric
redshifts because of the variable accuracy of the method due to
inhomogeneous exposure times in the five bands for the different
fields.\\
First, we use a color-color selection to isolate galaxies with $z\ge
0.7$, based on VVDS photometric redshifts estimation, relying
  on multi-color data \citep{Ilbert_2006}. As shown in
figure~\ref{fig_color_sel}, the $(g-r)$ versus $(r-i)$ selection
criterion is efficient to separate galaxies at $z\ge 0.7$ (big dots)
from the lower redshift population (small dots). The line shows our
separation criterion. 96\% of galaxies with $z_{phot}\ge 0.7$ are
located below the line while less than 10\% of low~$z$ objects
($z_{phot}\le 0.7$) fall in the same region.  \\
%

The photometric redshifts are computed by using {\it Le Phare} code
and by adopting the method described by \citet{Ilbert_2006}. The
comparison with the spectroscopic redshifts, obtained by the VVDS in
the best photometric field (CFHTLS-D1, \citet{Lefevre_2005}), for our
$u'$ selected sample shows an accuracy of $\sigma(\Delta
z/(1+z))=0.03$ with 4\% of outliers (defined as $\Delta z\ge
0.15\times (1+z)$).

\begin{figure}[t]
  \plotone{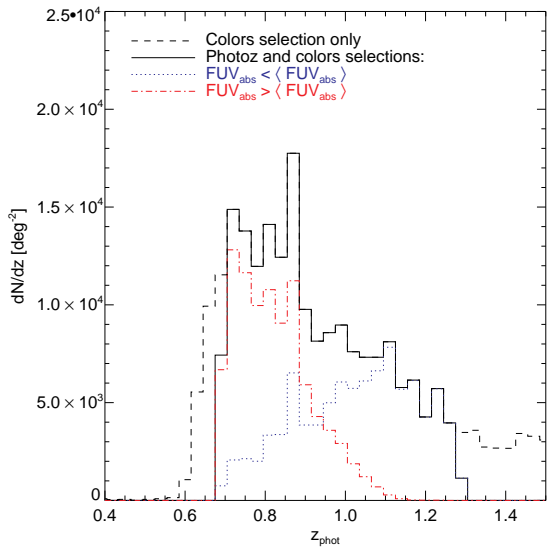}
  \caption{\small Redshift distribution of the CFHTLS sample: the
    dashed histogram shows the photometric redshift distribution of
    the galaxies selected with the color criterion alone, while the
    solid histogram shows the final redshift distribution after
    selecting objects with $0.7<z_{phot}<1.3$. The dotted and
    dot-dashed histograms show the redshift distributions of the
    $FUV_{\rm{abs}} < -19.41$ and $FUV_{\rm{abs}} > -19.41$
    respectively.}
  \label{fig_nz} 
\end{figure}

In figure~\ref{fig_nz} we show the photometric redshift distribution
of the galaxies selected with the color criterion (dashed
histogram). The final sample is obtained by further selecting objects
with $0.7<z_{phot}<1.3$ (solid histogram). The absolute magnitudes in
the \galex bands are derived from the best fitting SEDs whose NUV-rest
flux are well constrained by the $u'$, $g'$, and $r'$ bands in the
redshifts range ($0.7\le z\le 1.3$). Note that as the $u'$ filter
shifts to \FUV wavelengths at $z\sim 1$, absolute magnitudes depend
very weakly on k-correction and best-fit fitting SEDs. As for \galex
samples, we divide the CFHTLS sample in two bins according to the mean
\FUV absolute magnitude and the resulting redshift distributions are
shown in Fig~\ref{fig_nz}.\\
The global properties of the CFHTLS UV samples are given in table
\ref{tab_Muv_cfhtls}.

\section{Redshift evolution of the correlation function of UV-selected galaxies}\label{sec_results_Muv}
We compute the ACF using the \citet{Landy_1993} estimator. We assume
that the ACF is well approximated by a power-law: $w(\theta) =
A_{w}\theta^{-\delta}$; we use a variable Integral Constraint (IC)
with $\delta$ as free parameter during the power-law fitting process,
and estimate the IC with the same method used by
\citet{Roche_1999}. We derive correlation lengths ($r_0$) for each
sample from the Limber equation \citep{Peebles_1980}, using the
corresponding redshift distribution. These quantities, as well as the
bias parameter\footnote{See Paper I for details on the computations.},
are summarized in table~\ref{tab_Muv_galex} and
table~\ref{tab_Muv_cfhtls}. The effects of the dust internal to
galaxies have again been neglected.

\begin{figure*}[!t]
  \includegraphics[width=\hsize]{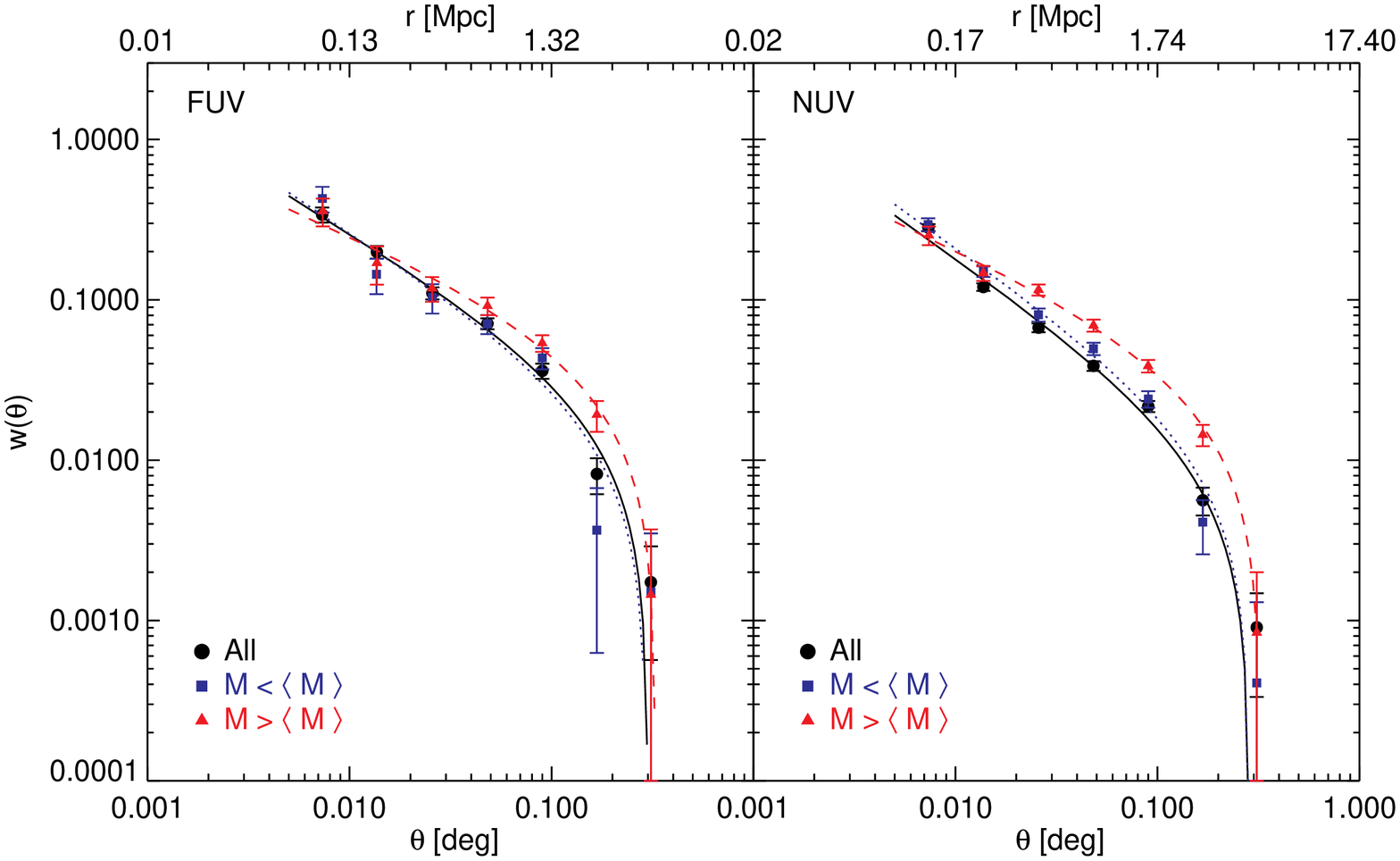}
  \caption{\small Angular correlation function of the \galex
  subsamples cut in absolute UV magnitude: $M<\langle M \rangle$,
  squares; $M>\langle M \rangle$, triangles and for comparison the
  total sample (circles). Left panel, \FUVnb; right panel, \NUVnb. The
  curves show the best fit power-law not corrected for the Integral
  Constraint bias. The upper axis shows the comoving scales
  corresponding to angular scales at $z = 0.18$ (left) or $z = 0.24$
  (right).}
  \label{fig_Muv_wtheta_galex}
\end{figure*}

\begin{figure}[!t]
  \plotone{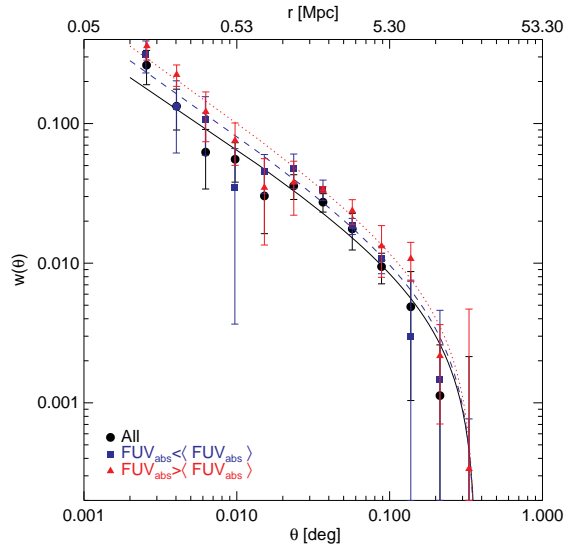}
  \caption{\small Angular correlation function of the CFHTLS
  subsamples cut in absolute UV magnitude: $M<\langle M \rangle$,
  squares; $M>\langle M \rangle$, triangles and for comparison the
  total sample (circles). The curves show the best fit power-laws with
  the Integral Constraint correction terms subtracted. The upper axis
  shows the comoving scales corresponding to angular scales at $z =
  0.9$.}
  \label{fig_Muv_wtheta_cfhtls}
\end{figure}

In figures~\ref{fig_Muv_wtheta_galex} and~\ref{fig_Muv_wtheta_cfhtls},
we show the ACFs of the \galex and CFHTLS samples respectively. The
ACFs are derived for the global samples and for two sub-samples with
UV absolute luminosity brighter and fainter than the mean
$<UV_{\rm{abs}}>$ of each sample. The angular scales probed for the
\galex samples are $0.005^{\circ}$ to $0.4^{\circ}$ (corresponding
respectively to comoving distances $0.07$ Mpc and $5.7$ Mpc at $z =
0.2$), while $0.002^{\circ}$ to $0.4^{\circ}$ for the CFHTLS samples
(resp. $0.11$ Mpc and $23$ Mpc at $z =1$). These ACFs are fairly well
fitted by power-laws, even if a small dip appears at small scales in
the FUV \galex samples and also in the CFHTLS bright one (see sec.
\ref{Halo_terms}). The higher surface density of UV-selected galaxies
at $z\sim 1$ allows a less noisy estimation of the ACF at these epochs
than at $z<0.4$. 

\begin{deluxetable}{cccccccc}
\tablecolumns{6} \tabletypesize{\scriptsize} 
\tablecaption{\small \galex samples description, power-law best fits parameters, comoving correlation lengths and bias\label{tab_Muv_galex}. }
\tablehead{
\colhead{}    &  \multicolumn{3}{c}{FUV samples} &   \colhead{}   &
\multicolumn{3}{c}{NUV samples} \\
\cline{2-4} \cline{6-8} \\ 
\colhead{} & \colhead{All} & \colhead{$FUV_{\rm{abs}}<-18.3$} & \colhead{$FUV_{\rm{abs}}>-18.3$} &
\colhead{}    & \colhead{All}   & \colhead{$NUV_{\rm{abs}}<-18.8$}    & \colhead{$NUV_{\rm{abs}}>-18.8$}  }
\startdata
$N_{gal}$\tablenotemark{*} & 42065 & 22082 & 19983 & & 97038 & 52567 & 44471\\[0.1cm]
$\langle FUV_{\rm{abs}} \rangle$ & -18.3 & -18.96 & -17.57 & & -18.23 & -18.76 & -17.61\\[0.1cm]
$\sigma_{FUV_{\rm{abs}}}$ & 0.91 & 0.52 & 0.67 & & 0.96 & 0.72 & 0.82\\[0.1cm]
$\langle NUV_{\rm{abs}} \rangle$ & -18.58 & -19.15 & -17.95 & & -18.8 & -19.43 & -18.05\\[0.1cm]
$\sigma_{NUV_{\rm{abs}}}$ & 0.84 & 0.52 & 0.67 & & 0.91 & 0.44 & 0.69\\[0.1cm]
$\langle z \rangle$\tablenotemark{\dagger} & 0.18 & 0.23 & 0.13 & & 0.25 & 0.32 & 0.17\\[0.1cm]
$\sigma_z$\tablenotemark{\dagger} & 0.08 & 0.07 & 0.05 & & 0.11 & 0.09 & 0.06\\[0.1cm]
$n_{gal}$ [$10^{-2}$Mpc$^{-3}$] &2.78$\pm$1.03 & 0.15$\pm$0.03 &2.54$\pm$0.95 & &2.16$\pm$1.19 &0.14 $\pm$0.03 & 1.95$\pm$ 1.06\\[0.1cm]
$A_{w}\times10^3$ & 9.4$^{+2.4}_{-1.7}$ & 7.3$^{+3.4}_{-2.2}$ & 33.9$^{+17.4}_{-10.3}$  & & 3.6$^{+0.6}_{-0.5}$ &4.1$^{+1.0}_{-0.8}$ & 22.4$^{+5.7}_{-4.6}$  \\[0.1cm]
$\delta$      & 0.74$\pm$0.05 & 0.79$\pm0.1$ & 0.48$\pm0.09$ & & 0.86$\pm0.04$ & 0.87$\pm0.05$ & 0.52$\pm0.05$   \\[0.1cm]
$r_0$ [Mpc]   & 4.6$^{+0.6}_{-0.5}$ & 4.6$^{+0.9}_{-0.7}$ & 5.4$^{+1.5}_{-1.0}$  & & 4.1$^{+0.3}_{-0.3}$ & 4.9$^{+0.4}_{-0.4}$& 5.5$^{+0.8}_{-0.7}$ \\[0.1cm]
$b_8$         & 0.74$^{+0.08}_{-0.07}$ & 0.76$^{+0.13}_{-0.1}$ & 0.83$^{+0.17}_{-0.12}$ & & 0.69$^{+0.05}_{-0.05}$ & 0.83$^{+0.07}_{-0.07}$ & 0.86$^{+0.09}_{-0.07}$  \\[0.1cm]
\enddata

\tablenotetext{*}{Number of galaxies in the samples}
\tablenotetext{\dagger}{according to photometric redshifts} \tablecomments{The
amplitude and slope of best fit power laws to the angular correlation
function, and hence the comoving correlation length account for the
Integral Constraint correction, as described in Paper I.}
\end{deluxetable}

\begin{deluxetable}{c c c c}
\tablecolumns{3} \tabletypesize{\footnotesize} \tablewidth{0pt}
\tablecaption{\small CFHTLS samples description, power-law best fits parameters, comoving correlation lengths and bias\label{tab_Muv_cfhtls}}
\tablehead{
  \colhead{} & \colhead{All} & \colhead{$FUV_{\rm{abs}}<-19.41$} & \colhead{$FUV_{\rm{abs}}>-19.41$} }
\startdata
$N_{gal}$\tablenotemark{*} & 17098  & 8507   & 8591\\[0.1cm]
$\langle FUV_{\rm{abs}}\rangle $ & -19.41 & -19.89 & -18.94    \\[0.1cm]
$\sigma_{FUV_{\rm{abs}}}$ & 0.6  & 0.34   & 0.36 \\[0.1cm]
$\langle NUV_{\rm{abs}}\rangle $ & -19.81 & -20.19 & -19.43    \\[0.1cm]
$\sigma_{NUV_{\rm{abs}}}$ & 0.53  & 0.4   & 0.33 \\[0.1cm]
$\langle z\rangle $\tablenotemark{\dagger} & 0.94   & 1.04   & 0.84  \\[0.1cm]
$\sigma_z$\tablenotemark{\dagger} & 0.16 & 0.15 & 0.09\\[0.1cm]
$n_{gal}$ [$10^{-3}$Mpc$^{-3}$] &3.27$\pm$2.90 &0.62$\pm$0.41 & 2.66$\pm$2.58\\[0.1cm]
$A_{w}\times10^3$ & $2.7^{+1.4}_{-1.0}$ & $2.8^{+1.2}_{-0.8}$ & $3.3^{+1.0}_{-0.7}$ \\[0.1cm]
$\delta$      & $0.7\pm0.09$     & $0.74\pm0.09$   & $0.76\pm0.07$\\[0.1cm]
$r_0$ [Mpc]   & $4.92^{+0.5}_{-0.5}$& $5.48^{+0.5}_{-0.5}$ &$4.66^{+0.24}_{-0.23}$ \\[0.1cm]
$b_8$         & 1.24$^{+0.08}_{-0.07}$ & 1.38$^{+0.06}_{-0.07}$ & 1.16$^{+0.03}_{-0.04}$ 
\enddata

\tablenotetext{*}{Number of galaxies in the samples}
\tablenotetext{\dagger}{according to photometric redshifts} \tablecomments{Same note as in table \ref{tab_Muv_galex}}

\end{deluxetable}

\subsection{Clustering segregation with \FUV luminosity }
\label{sec_ro_Muv}

\begin{figure}[!t]

  \plotone{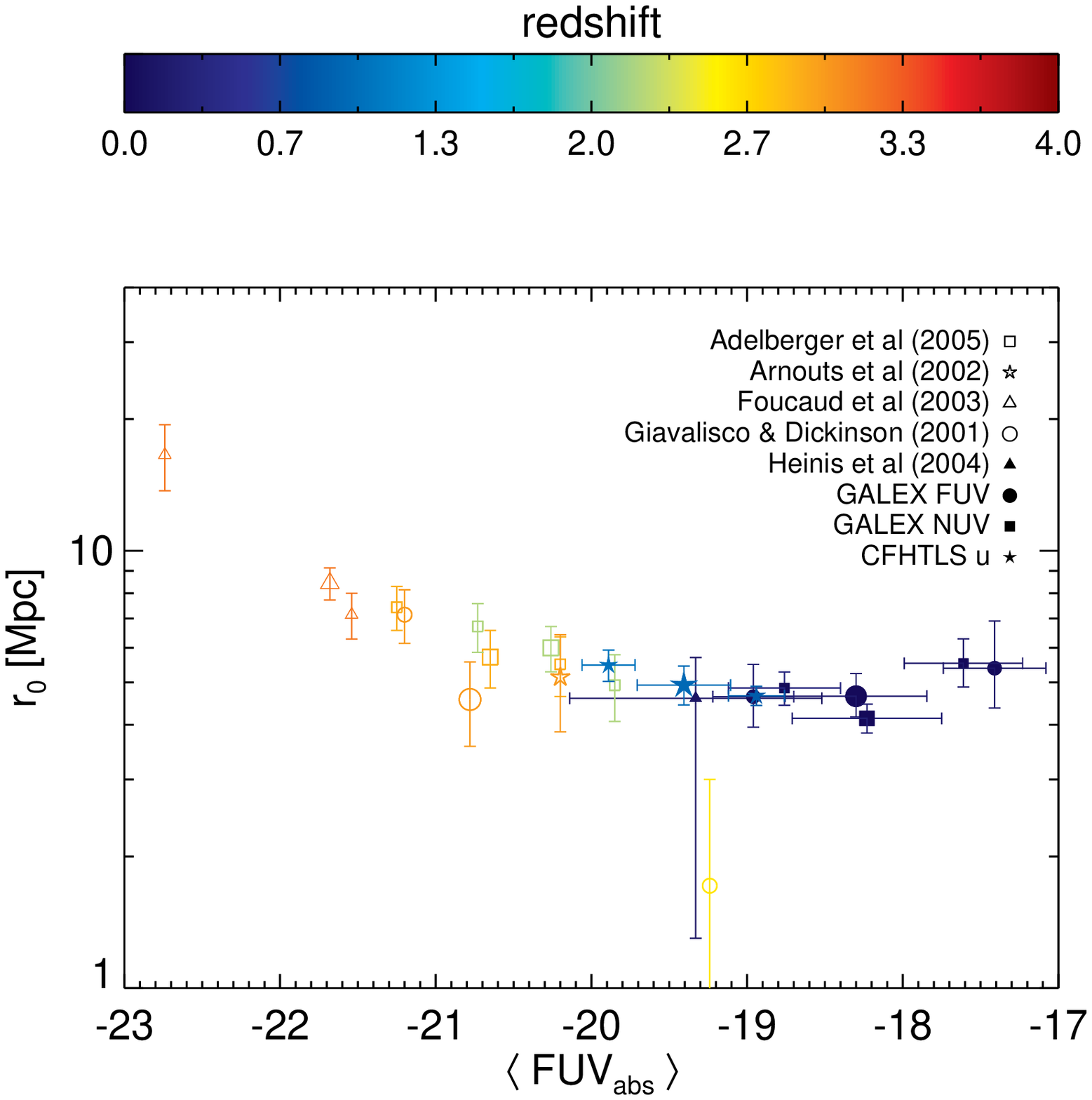}
  \caption{\small Dependence on absolute \FUV magnitude of the
  correlation length $r_0$ for low and high redshift restframe
  $UV$-selected galaxies: open squares, \citet{Adelberger_2005}; open
  star, \citet{Arnouts_2002}; open triangles, \citet{Foucaud_2003};
  open circles, \citet{Giavalisco_2001}. Our results are presented as
  filled circles (\FUVnb) and filled squares (\NUVnb) for the \galex
  samples and as filled stars for the CFHTLS samples. The mean
  redshifts of the samples are color-coded. Note that the results from
  a given study (including ours) are not all obtained from independent
  samples. Hence we distinguish global samples by plotting them with a
  bigger symbol size, in this figure and in the following ones as
  well, unless otherwise stated.}
  \label{fig_ro_Muv}
\end{figure}

The dependence of $r_0$ on \FUV luminosity in \galex and CFHTLS
samples is shown in fig.~\ref{fig_ro_Muv}, along with results from higher
redshift studies ($z\ge 2$)\footnote{We choose the $FUV$ absolute
magnitude for the comparison as most of high redshift samples are
$FUV$ restframe selected, and \galex results are not strongly
dependent of the UV band.\\ The mean absolute magnitudes of the LBG
samples have been obtained by deriving an average apparent magnitude
from the galaxy counts, and assuming a k-correction of
$2.5\log(1+z)$. \citet{Ouchi_2005} do not provide their counts, so we
computed the expected mean absolute magnitude given their limiting
absolute magnitude and the luminosity function of
\citet{Sawicki_2006}.}. As the different surveys probe different parts
of the UV luminosity function with little overlap, it is difficult to
draw firm conclusions. Nevertheless, significant differences between
the samples are apparent:

\begin{itemize}    
\item We use as reference the correlation function results from
\citet{Adelberger_2005}, \citet{Allen_2005}, \citet{Foucaud_2003},
\citet{Giavalisco_2001}, \citet{Lee_2006} and \citet{Ouchi_2001} at
$z>2$. At these redshifts, all studies conclude a significant
segregation of $r_0$ with UV luminosity (the more luminous the more
clustered) in the range $-23 \le FUV_{\rm{abs}}\le -20$.
\item At $z\sim 1$, a positive correlation of $r_0$ with \FUV is still
observed for $-20\le FUV_{\rm{abs}} \le -19$. Notably, our value of
$r_0$ at $FUV_{\rm{abs}} \sim -20$ is very close to that of
\citet{Adelberger_2005} obtained from $z \sim 2$ samples.
\item At $z\lesssim0.3$, we probe a fainter luminosity range ($-19 \le
 FUV_{\rm{abs}} \le -17$), and a weak anti-correlation of $r_0$ with \FUV is
 apparent, though given the error bars, it is compatible with no \FUV
 luminosity segregation of $r_0$.
\end{itemize}

\begin{figure}[!t]
  \plotone{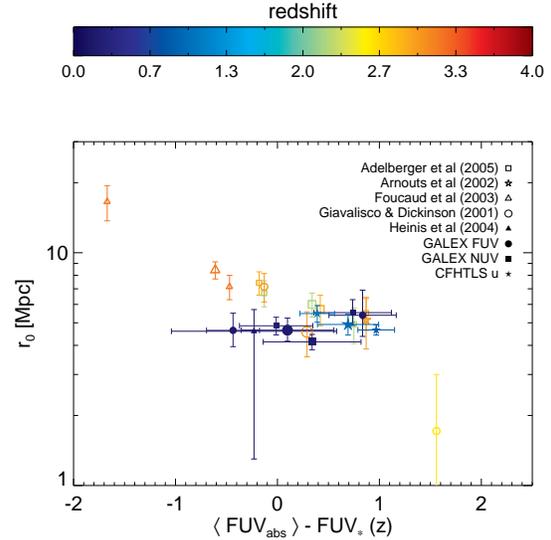}
  \caption{\small Same as in figure \ref{fig_ro_Muv} but versus $FUV_{\rm{abs}} -
  FUV_{\ast}$. }
  \label{fig_ro_Mstar}
\end{figure}

The values of $r_0$ as a function of \FUV luminosity for \FUV selected
samples at different redshifts follow a unique smooth curve, with a
significant slope at the bright end $FUV_{\rm{abs}} \le -19$ and a
flat or slightly negative slope at the faint end. A similar
segregation is observed with $B$ luminosity at low redshift, with
optical selection criteria \citep{Benoist_1996, Guzzo_2000,
Norberg_2002, Willmer_1998, Zehavi_2005}. In particular,
\citet{Norberg_2001} showed that for blue-selected galaxies $r_0$
increases only slowly for galaxies fainter than $L^{B}_{*}$, while it
varies strongly for galaxies brighter than $L^{B}_{*}$. Indeed, using
$N$-body simulations \citet{Benson_2001} showed that $L^{B}_*$ could
be a natural boundary in the distribution of the halos hosting
galaxies, galaxies fainter than $L^{B}_*$ being hosted by a
mix of low and high mass halos, while galaxies brighter than $L^{B}_*$
hosted by more and more massive haloes. To check if $FUV_*$ could
play a similar role in UV samples, we show in
figure~\ref{fig_ro_Mstar} $r_0$ as a function of $\langle
FUV_{\rm{abs}} \rangle - FUV_{\ast}$, where the evolution of
$FUV_{\ast}$ with $z$ is taken from \citet{Arnouts_2005} (for $z<1$)
and \citet{Sawicki_2006} (for $z>1$). The luminosity dependence of
$r_0$ changes noticeably when expressed as a function of
$FUV_{\rm{abs}}-FUV_{\ast}$, as two different trends are observed
according to the redshift range:

\begin{itemize}
 \item At $z\ge 1$, for the high $z$ samples and our CFHTLS sample,
 the behavior of $r_0$ with $FUV_{\rm{abs}}-FUV_{\ast}$ is
 qualitatively compatible with the monotonic trend described
 above, the brighter galaxies being more clustered.
 \item At $z\le0.5$ (\galex samples) a radically different behavior of
 $r_0$ vs $FUV_{\rm{abs}}-FUV_{\ast}$ is seen. An anti-correlation or
 no correlation (given the errorbars) is observed, with brighter
 samples showing slightly lower $r_0$ than fainter ones.
\end{itemize}
This suggests that the luminosity segregation mechanisms of the
clustering at low redshifts work in a different regime, or that
$FUV_{\ast}$ is not the relevant variable.

\subsection{ACF slope segregation with \FUV luminosity } 
\label{sec_delta_Muv}

\begin{figure}[!t]
  \plotone{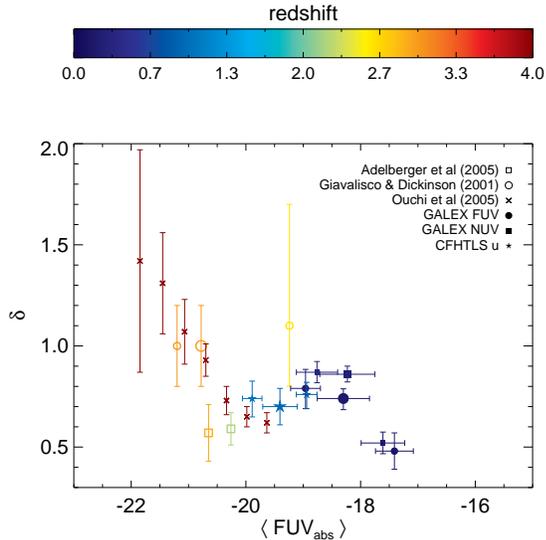}
  \caption{\small Dependence on absolute \FUV magnitude of the slope
  $\delta$ of the correlation function for low and high redshift
  restframe $UV$-selected galaxies. Results shown here come from
  studies allowing $\delta$ as a free parameter: open squares,
  \citet{Adelberger_2005}; open circles, \citet{Giavalisco_2001};
  crosses, \citet{Ouchi_2005} (results from the fit with Integral
  Constraint). Our results are presented as filled circles (\FUVnb)
  and filled squares (\NUVnb) for the \galex samples and as filled
  stars for the CFHTLS samples; the error on $\delta$ comes from the
  fitting procedure and the horizontal bars for our samples reflect
  the standard deviations of $FUV_{\rm{abs}}$. The mean redshifts of the
  samples are color-coded.}
  \label{fig_delta_Muv}
\end{figure}

\subsubsection{ACF slope}
The slope of the ACF ($\delta$), which describes the balance between
small and large scale separations, is an important indicator on the
nature of the spatial distribution of a given population. In paper I,
we found that the estimates of the slope inferred from the global
\galex samples ($\delta \simeq 0.81\pm 0.07$) are steeper than those
derived from optically selected blue galaxies in the local Universe:
$\delta\sim 0.6$ \citep{Budavari_2003, Madgwick_2002, Zehavi_2002,
Zehavi_2005}. \\
In figure~\ref{fig_delta_Muv}, we now analyse the dependence of
$\delta$ on UV luminosity for the different samples (\galex samples:
filled circles for \FUV and filled squares for \NUVnb; CFHTLS samples:
filled triangles; high $z$ samples: empty squares, open circles and
crosses; all points color-coded with redshift).\\

At $z > 3$, the compilation of measurements showed here, and
especially those at at $z = 4$, from \citet{Ouchi_2005} indicate that
the ACF slope steepens at higher UV luminosities.

\citet{Ouchi_2005} claimed that this trend, well modeled in the Halo
Occupation Distribution (HOD) framework, is not an actual slope
variation but it is related to the halo occupancy. Based on HOD
models, they show that the contribution of satellite galaxies ('one
halo term', see e.g. \citet{Zehavi_2004}) increases when selecting
brighter galaxies, by enhancing the small scale signal of the ACF ($ r
\lesssim 0.35 $ Mpc). This effect produces an apparent steepening of
the observed slope $\delta$.\\ Our GALEX sample at $z<0.4$ seems to
produce a similar although less pronounced effect. Our current low $z$
GALEX data do not allow to perform detailed comparison between
observations and HOD models, but we have investigated if the observed
steepening with luminosity can be partially due to the small scale
component. We fitted the GALEX ACF only at scales $r>0.4$ Mpc (see
sec. \ref{Halo_terms}) or $r>0.7$ Mpc in order not to include the one
halo term component. We do not observe any significant change with
respect to our initial slopes. However doing so we face at large
scales the problems of lower signal-to-noise ratio and more important
contribution of the Integral Constraint bias that prevent us to make
firm statements. This test thus relies on the efficiency of our power
law fitting process in recovering the true ACF (see Paper I).\\In
other words, at low redshift, we do not see evidence that the one halo
term plays a major role in the slope of the ACF, as observed at high
redshift, which is expected from simulations
\citep{Kravtsov_2004}. Hence this indicates that our clustering
parameters ($r_0$, $\delta$ and bias, $b_8$) reflect the large scale
clustering of star forming galaxies, which enables us to make
comparisons with analytical predictions for the clustering of Dark
Matter Haloes.

\subsubsection{Dip in the ACF ?}\label{Halo_terms}

The ACFs derived for the various \galex and CFHTLS samples are
globally well described by a power-law, but some of our ACFs show a
little dip around 0.35 Mpc, the \galex \FUV ones for instance, and
also the brightest CFHTLS sample at $z\sim1$ (at a slightly larger
scale $\sim 0.5$ Mpc). This recalls the departure to the power-law
observed in other surveys and interpreted as the transition between
the one and two halo terms in the HOD framework. \citet{Zehavi_2004}
showed that this transition occurs at $\sim 1.5-3$ Mpc for $r$-band
selected galaxies. It is expected that this scale should be shorter
for bluer galaxies, residing in less massive halos, as showed by
\citet{Magliocchetti_2003} in observations and \citet[][see their
fig. 22]{Berlind_2003} in simulations, with a transition scale for
late-types galaxies at $\sim0.45$ Mpc, close to what we
observe. Finally and very interestingly \citet{Ouchi_2005} observe
this transition for LBGs at $z\sim4$ at 0.35 Mpc, the same comoving
scale as we get.

Comparing measurements with predictions from HOD models is a natural
perspective of this work, to probe the redshift evolution of the halo
occupancy of star-forming galaxies. This will be addressed in details
in a forthcoming paper with enlarged datasets.

\section{Bias of star-forming galaxies from $z = 4$ to $z = 0 $}
\label{sec_bias}

The link between the properties of the galaxy distribution and the
underlying Dark Matter density field can be accessed via the bias
formalism. The bias parameter is indicative of the masses of the dark
matter halos that preferentially host the observed galaxy population
\citep[e.g.][]{Giavalisco_2001, Mo_2002, Ouchi_2004}, {\it i.e.} in
our case, actively star forming galaxies. The DM halo bias is a direct
output of \citet{Mo_2002} models. For galaxies, we assume a linear
bias to convert $r_0$ in $\sigma_{8,g}$, a common, though
questionable, assumption \citep[see e.g.][]{Marinoni_2005}.

\begin{figure}[htbp]
  \plotone{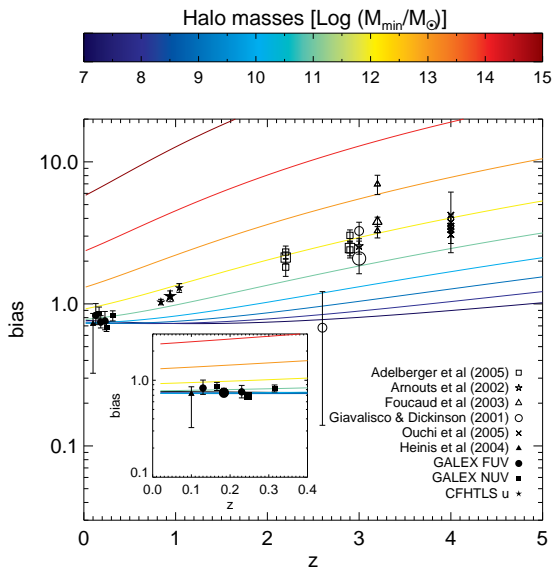}
  \caption{\small Evolution with redshift of the bias of restframe
   $UV$-selected galaxies (symbols as figure \ref{fig_ro_Muv}). The
   curves show the effective bias of halos more massive than $M_{min}$
   (color-coded) according to \citet{Mo_2002} (see also Paper I). The
   inset is an enlargement of the low redshift area.}
  \label{fig_bias_z}
\end{figure}

\subsection{Redshift evolution of the bias}
Figure~\ref{fig_bias_z} shows as symbols the redshift evolution of the
bias parameter measured at 8 $h^{-1}$ Mpc defined as $b_8 =
\sigma_{8,g} / \sigma_{8,m}$ \citep[see e.g.][]{Magliocchetti_2000}
for the different samples discussed above.\footnote{For
\citet{Adelberger_2005} subsamples, $\delta$ values are not available;
we assumed that the slopes are the same than those of their global
samples. The expected relative error on the inferred bias is lower
than 10\% if $0<\delta<1$.} The bias values for our \galex and CFHTLS
samples are reported in tables \ref{tab_Muv_galex} and
\ref{tab_Muv_cfhtls} respectively. \\
The observed bias of star-forming galaxies shows a gradual increase
with look back time:
at $z>2$, UV galaxies are strongly biased \citep{Giavalisco_2001,
Foucaud_2003}, with $b_8\gtrsim 2$, and at a given redshift the bias
increases with \FUV luminosity (\FUV luminosity segregation).
At $z\sim 1$, the mean bias is $\langle b_8 \rangle =1.26 \pm0 .06$,
indicating that star-forming galaxies are closer tracers of the
underlying mass distribution at that time.
At $z\le 0.4$, given the error bars, the mean bias is consistent with
0.8 for all \galex samples ($\langle b_8 \rangle =
0.79^{+0.1}_{-0.08}$), a slight anti-bias independent of the UV
luminosity.\\

In figure~\ref{fig_bias_z}, we also show the effective bias evolution
derived from the \citet{Mo_2002} formalism for different minimum Dark
Matter Halo (DMH) mass thresholds. A comparison can be made to the
bias of star forming galaxies, if one assumes that most haloes do not
host more than one star-forming galaxy. This coarse assumption is
likely inaccurate for star-forming galaxies selected at high redshifts
in \FUV with a well developed one-halo term \citep{Kashikawa_2006,
Lee_2006}, but is acceptable at low redshifts in the \FUV since the
one-halo term does not seem to play a major role as discussed
sect. \ref{sec_delta_Muv}. \\

The mimimum masses of the DMH that produce the bias derived for
galaxies are $10^{12} M_{\odot} \lesssim M \lesssim 10^{13} M_{\odot}$
at $z\ge 2$, $10^{11} M_{\odot} \lesssim M \lesssim 10^{12} M_{\odot}$
at $z\simeq 1$ and $M \le 10^{12} M_{\odot}$ at $z<0.4$. There is an
obvious degeneracy of the models at low redshifts, but the locally
observed bias is definitely in the region of low cutoff masses. This
is a hint that observed star-forming galaxies at low redshift reside
preferentially in less massive halos than high $z$ star-forming
galaxies.

\begin{figure}[htbp]
  \plotone{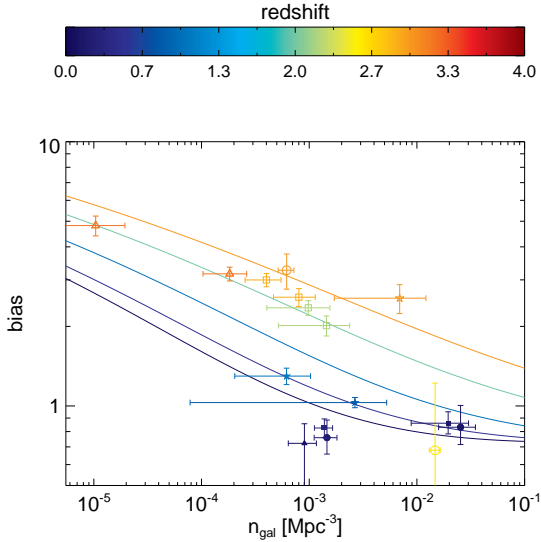}
  \caption{\small Bias of restframe $UV$-selected galaxies as a
   function of galaxy number density. The curves show the expected
   relation for the effective bias of Dark Matter halos at redshifts
   $z=0,0.5,1,2,3$, from bottom to top. We plot only independent
   samples here. The legend is the same as in fig.~\ref{fig_bias_z}.}
  \label{fig_bias_ngal}
\end{figure}

\subsection{Bias and galaxy number density}
In fig.~\ref{fig_bias_ngal} we show the bias a function of the galaxy
number density $n_{gal}$ for UV-selected samples and the predicted
relation between the effective bias and the number density of DMHs at
$z=0,0.5,1,2,3$ (curves from bottom to top). At high redshift ($z>1$),
we observe the well known luminosity segregation effect, brighter
galaxies (less abundant) having a larger bias, in good agreement with
DMH models predictions (the less abundant, the more clustered). In
contrast at low $z$ ($z<1$), a significant departure to this relation
is observed.  At $z\sim 0.9$, the CFHTLS data show a bias slightly
lower than the expected one according to the observed density with
$n_{gal}$ approximately 3 times lower than expected. This seems even
worse for our brightest samples at $z\sim 0.3$, as these galaxies are
about 10 times less numerous than expected according to their bias
values. In the model discussed here, we implicitly assume that one DMH
hosts one galaxy, which provides a fairly reasonable description of
the observations at high $z$, to the level of precision allowed here
\citep[see e.g.][for more detailed discussions on this
point]{Adelberger_2005, Ouchi_2004}. At $z<1$, this assumption may be
not valid anymore and our results suggest that star forming galaxies
(especially the brightest) are not hosted by a significant fraction of
the DMHs with similar clustering properties. This implies that the DMH
occupation fraction, that is roughly $>0.5$ at high redshift ($z>2$),
drops to 0.3 and 0.1 at $z=1$ and $z=0.3$ respectively.

%

\subsection{Bias and FUV LD fraction}
\begin{figure}[htbp]
  \plotone{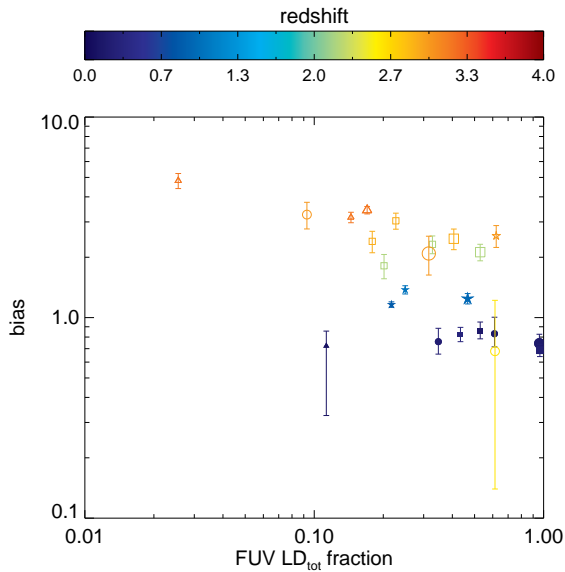}
  \caption{\small Bias of restframe $UV$-selected galaxies (symbols as
   in figure \ref{fig_bias_z}) as a function of the fraction of the
   FUV total luminosity density represented by each sample. The legend
   is the same as in fig.~\ref{fig_bias_z}.}
  \label{fig_bias_LD}
\end{figure}

The very limited overlap in \FUV luminosities of the data at different
redshifts does not allow a derivation of the bias evolution with
redshift at fixed \FUV luminosity. However, despite the fact that low
$z$ samples reach fainter luminosities than high $z$ ones, they happen
to probe the same fraction of total \FUV luminosity densities, owing
to the strong evolution of the \FUV luminosity function with $z$
\citep{Arnouts_2005}. In particular at all redshifts the samples are
able to probe the bulk of star formation, {\it i.e.}  they encompass a
fraction of the \FUV luminosity density (LD) greater than 0.5. This
can be seen in fig.~\ref{fig_bias_LD} where we show the bias as a
function of the fraction of the total \FUV LD enclosed by the
different samples. This favorable situation allows us to track the
evolution with redshift of the clustering at a fixed fraction of the
\FUV LD, an essentially constant fraction of the star formation rate.

The fraction of the LD for each sample is computed by comparing the
total LD at the relevant redshift (from the \FUV Luminosity function
parameters of \citet{Arnouts_2005} and \citet{Sawicki_2006}), to the
LD enclosed by each sample according to its flux limits converted to
\FUV luminosity cuts\footnote{Note that while LBG samples are by
construction volume-limited, we cannot adopt this approach for the
\galex samples due to limited statistics. This means that faintest
galaxies are underrepresented, especially in our higher redshift
\galex samples. However, as we do not observe a strong luminosity
dependence of the bias within the \galex samples, we expect this has
only a small impact.}.\\ Note that we do not attempt to correct for
galaxy internal dust attenuation. As brighter UV galaxies present
higher extinction in the local Universe \citep{Buat_2005}, the strong
brightening of $FUV_{\ast}$ with redshift \citep[e.g.][]{Arnouts_2005}
may introduce a small bias as a given LD fraction could not correspond
exactly to the same star formation rate fraction at the different
redshifts we explore here.

The plot confirms the result already apparent in fig.~\ref{fig_bias_z}
of a significant decrease of the bias of UV selected galaxies from
high to low redshifts, but now selected on the basis of a physically
defined parameter, the fraction of the \FUV luminosity density. Near
an LD fraction of $0.5$, the bias is divided by a factor 3, between
redshifts near 3, shifting from $2.5$ in the redshift range $2-3$ down
to $0.8$ in the local universe.

\section{Discussion}\label{sec_discussion}

In paper I we reported on the overall clustering properties of the
UV-selected galaxies in \galex samples, the largest ones available to
date at low redshift and at these wavelengths. These samples allow for
the first time an investigation of the clustering properties of
UV-selected galaxies as a function of different parameters at
$z\lesssim 1$, which can be compared to higher redshift samples also
selected in the rest frame FUV.

The measurements from the \galex samples confirm previous results for
rest-UV selected galaxies at low redshifts indicating that they are
weakly clustered \citep{Heinis_2004}, with an autocorrelation function
well approximated by a power law in the range $0.2-5$ Mpc.

At $z \sim 1$, the correlation length of the rest-UV selected galaxies
from CFHTLS $u'$ data is found comparable to those of the
emission-line samples from \citet{Coil_2004} in the same redshift
range, but slightly higher than those obtained by \citet{Meneux_2006}
for late type and irregular galaxies (their types 3 and 4, selected in
the visible). At those redshifts, according to our CFHTLS sample, star
forming galaxies are modestly biased with $\langle b_8 \rangle =1.26
\pm 0.06$, which under the linear bias hypothesis implies they are
closer tracers of the mass distribution than their higher redshift
counterparts. As opposed to the dependence found at redshifts above
$2$ \citep{Giavalisco_2001, Foucaud_2003}, no strong positive
correlation between the bias and the \FUV luminosity is observed in
the local universe, but rather a slight anti-correlation or no
correlation. At $z\le 0.4$, given the error bars, the mean bias is
consistent with 0.8 for all \galex samples ($\langle b_8 \rangle =
0.79^{+0.10}_{-0.08}$) independently of the UV luminosity.

\subsection{\it Migration of the bulk of star formation sites from $z = 3$ to the local universe}

In this study, we find a decrease by a factor 3.1 of the bias with
respect to mass, from redshifts near $3$ to the local universe in the
UV flux-limited samples, and more importantly in samples selected in
UV luminosity so that they encompass a constant fraction of the
luminosity density at all $z$. This decrease is slightly larger than
the factor 2.7 derived from the \citet{Mo_2002} model for the $M \ge
10^{12} M_{\odot}$ haloes that host most star formation at redshift
$3$, an indication that star forming galaxies tend to be hosted by
haloes of lower mass in the local universe. This is the main
conclusion of the present study.

The ``downsizing'' scenario \citep{Cowie_1996, Juneau_2005,
Bundy_2005, Heavens_2004} states that the star formation shifts from
high stellar mass systems at high redshift to low ones at low
redshift. Our results extend this vision in the sense that the same
trend is observed for the mass of the dark matter halos that host
actively star forming galaxies.

The DMH mass migration of the bulk of the star formation might be
associated with regions of different densities. At high redshifts, LBGs
studies show that active star formation traced by the UV light resides
preferentially in overdense regions \citep{Adelberger_1998,
Blaizot_2004, Giavalisco_2002, Steidel_1998, Tasker_2006}. At low
redshift, \citet{Abbas_2005} showed that the slope of the fitted power
law is steeper in underdense regions, and that the correlation length
is smaller. The observed steeper ACFs for the more UV-luminous
galaxies at low $z$ suggest that the most star-forming objects reside
preferentially in regions where the local galaxy density is lower than
for the fainter ones, a result in agreement with direct optical based
studies of star formation as a function of galaxy density in the local
universe \citep{Gomez_2003, Lewis_2002}.

\acknowledgments
We thank Christian Marinoni for stimulating discussions.

\galex (Galaxy Evolution Explorer) is a NASA Small Explorer, launched
in April 2003. We gratefully acknowledge NASA's support for
construction, operation, and science analysis for the \galex mission,
developed in cooperation with the Centre National d'Etudes Spatiales
of France and the Korean Ministry of Science and Technology.

This study is also based on observations obtained with
MegaPrime/MegaCam, a joint project of CFHT and CEA/DAPNIA, at the
Canada-France-Hawaii Telescope (CFHT) which is operated by the
National Research Council (NRC) of Canada, the Institut National des
Science de l'Univers of the Centre National de la Recherche
Scientifique (CNRS) of France, and the University of Hawaii. This work
is based in part on data products produced at TERAPIX and the Canadian
Astronomy Data Centre as part of the Canada-France-Hawaii Telescope
Legacy Survey, a collaborative project of NRC and CNRS.

\end{document}